\documentclass[12pt]{article}

\catcode`@=12
\begin{document}
\thispagestyle{empty}
\begin{flushright}
hep-ph/0412018\\ 
UCRHEP-T382\\ 
December 2004\
\end{flushright}
\vspace{0.5in}
\begin{center}
{\LARGE	\bf  Smallness of Leptonic $\theta_{13}$\\ and Discrete Symmetry\\}
\vspace{1.0in}
{\bf Shao-Long Chen, Michele Frigerio, and Ernest Ma\\}
\vspace{0.2in}
{\sl Physics Department, University of California, Riverside, 
California 92521\\}
\vspace{1.0in}
\end{center}
\begin{abstract}\
The leptonic mixing angle $\theta_{13}$ is known to be small.  If it is 
indeed tiny, the simplest explanation is that charged leptons mix only in the
$\mu-\tau$ sector and neutrinos only in the $1-2$ sector. 
We show that this 
pattern may be explained by the discrete symmetry $Z_2 \times 
Z_2$ of a complete Lagrangian, which has 2 Higgs doublets and 2 Higgs 
triplets (or 2 heavy right-handed singlet neutrinos). In the case of Higgs 
triplets, the
Majorana neutrino masses are arbitrary, whereas in the case of heavy singlet 
neutrinos, an inverted hierarchy is predicted.  
Lepton-Flavor-Violation effects, present only in the $\mu-\tau$ sector, 
are analyzed in detail: the LFV $\tau$-decay rates are predicted 
below the present bounds by a few orders of
magnitude, whereas LFV Higgs decays could allow
for a direct test of the model.

\end{abstract}    

\begin{center}
\end{center}

\newpage
\baselineskip 24pt

Recent experimental advances in measuring the neutrino oscillation parameters 
in atmospheric and solar data \cite{review} have now fixed the $3 \times 3$ 
lepton mixing matrix $U$ to a large extent.  Assuming that the neutrino mass 
matrix ${\cal M}_\nu$ is Majorana and it  is written in the basis where 
the charged-lepton mass matrix is diagonal, then for
\begin{equation}
U^T {\cal M}_\nu U = \pmatrix {m_1 & 0 & 0 \cr 0 & m_2 & 0 \cr 0 & 0 & 
m_3}
\label{one}\end{equation}
with the convention
\begin{equation}
U= \pmatrix {1 & 0 & 0 \cr 0 & c_{23} & -s_{23} \cr 0 & s_{23} & c_{23}} 
\pmatrix {c_{13} & 0 & s_{13} e^{-i\delta} \cr 0 & 1 & 0 \cr -s_{13} 
e^{i\delta} & 0 & c_{13}} \pmatrix {c_{12} & -s_{12} & 0 \cr s_{12} & c_{12} 
& 0 \cr 0 & 0 & 1},
\label{U}
\end{equation}
present data imply that $\theta_{23}$ is close to $\pi/4$, $\theta_{12}$ is 
large but far from $\pi/4$, and $\theta_{13}$ is small and consistent with 
zero ($\sin^2\theta_{13}\le0.047$ at $3\sigma$ C.L. \cite{valencia}). 

If data will significantly strengthen the upper bound on $\theta_{13}$,
this will imply a very special pattern for the violation of lepton flavor, 
which begs for a theoretical rationale.  In fact, it is possible to 
define quantitatively and experimentally when the $1-3$ mixing can be 
considered negligible: a value
as tiny as $\sin^2\theta_{13} \le 10^{-4}$ can be generated by gravity effects 
alone \cite{vibe} and neutrino factories could be sensitive to such small 
mixing \cite{nufa}.

The question of whether the origin of the lepton mixing $U=U_l^\dag U_\nu$ 
is in the neutrino or the charged-lepton sector has been discussed in many 
recent papers, e.g. \cite{frpe,alfe,roma,anki,raid,feru}. Just from the form of
Eq.~(\ref{U}), it is apparent which is the most simple-minded realization of 
zero $1-3$ mixing: besides the diagonal contributions to the neutrino and 
charged-lepton mass matrices ${\cal M}_\nu$ and ${\cal M}_l$, 
one needs to generate off-diagonal 
entries only in the $1-2$ sector of ${\cal M}_\nu$ and in the $\mu-\tau$ 
sector of 
${\cal M}_l$.  In this case the atmospheric mixing 
originates in the charged-lepton 
sector and the solar mixing in the neutrino sector.  In particular, this 
hybrid scenario has been shown to be generically associated with small 
values of $\theta_{13}$ \cite{mism}.

We point out in this paper that the above-mentioned hybrid scenario with 
$\theta_{13}=0$ is realized by a discrete symmetry of the Lagrangian 
of a complete theory, with distinct experimentally verifiable predictions.
Other models predicting $\theta_{13}=0$ have also been proposed 
\cite{grla,low,gjklt,gjklst}.

Consider the discrete symmetry $Z_2 \times Z_2$, also known as 
the Klein group. There are 4 possible 
representations, i.e. $(+,+)$, $(+,-)$, $(-,+)$, and $(-,-)$.  Suppose the 
3 lepton families transform as follows:
\begin{equation}
(\nu_i,l_i), ~l^c_i \sim (+,-), (-,+), (-,-),
\label{3nt1}
\end{equation}
with 2 Higgs doublets
\begin{equation}
(\phi_1^0,\phi_1^-) \sim (+,+), ~~~ (\phi_2^0,\phi_2^-) \sim (+,-),
\end{equation}
and 2 Higgs triplets
\begin{equation}
(\xi_1^{++},\xi_1^+,\xi_1^0) \sim (+,+), ~~~ (\xi_2^{++},\xi_2^+,\xi_2^0) 
\sim (-,-).
\end{equation}
Then the charged-lepton mass matrix linking $l_i$ to $l^c_j$ is given by
\begin{equation}
{\cal M}_l = \pmatrix{a & 0 & 0 \cr 0 & b & d \cr 0 & e & c},
\label{ml}\end{equation}
where the diagonal entries $a,b,c$ are induced by $\langle \phi_1^0 \rangle$, 
and $d,e$ by $\langle \phi_2^0 \rangle$, 
and the Majorana neutrino mass matrix is given by
\begin{equation}
{\cal M}_\nu = \pmatrix{A & D & 0 \cr D & B & 0 \cr 0 & 0 & C},
\label{mnu}\end{equation}
where $A,B,C$ come from $\langle \xi_1^0 \rangle$, and $D$ from $\langle 
\xi_2^0 \rangle$.  
The Higgs triplets are assumed to be very heavy ($\sim M_\xi$), 
so that they acquire naturally 
small vacuum expectation values $(\sim \langle \phi_i^0 \rangle^2 / M_\xi)$ 
\cite{ms98}.

Then ${\cal M}_l$ is diagonalized by a 
rotation in the $2-3$ sector and ${\cal M}_\nu$  by a 
rotation in the $1-2$ sector.  Hence $U$ is exactly of the form desired 
with $\theta_{13}=0$ (models predicting Eqs. (\ref{ml}) and (\ref{mnu}) 
by using different discrete symmetries can be found in \cite{low}).  
In particular,
\begin{eqnarray}
\pmatrix{a & 0 & 0 \cr 0 & b & d \cr 0 & e & c} &=& \pmatrix{1 & 0 & 0 \cr 
0 & c_L & s_L \cr 0 & -s_L & c_L} \pmatrix{m_e & 0 & 0 \cr 0 & m_\mu & 0 \cr 
0 & 0 & m_\tau} \pmatrix{1 & 0 & 0 \cr 0 & c_R & -s_R \cr 0 & s_R & c_R} 
\nonumber \\ &=& \pmatrix{m_e & 0 & 0 \cr 0 & c_L c_R m_\mu + s_L s_R m_\tau 
& -c_L s_R m_\mu + s_L c_R m_\tau \cr 0 & -s_L c_R m_\mu + c_L s_R m_\tau 
& s_L s_R m_\mu + c_L c_R m_\tau}, 
\label{defi}\end{eqnarray}
with $s_L = s_{23}$, $c_L = c_{23}$.  As for $\theta_{12}$, it is determined 
by Eq.~(7) which also allows for arbitrary $m_{1,2,3}$.  In other words, 
this model does not constrain any mass or mixing other than $\theta_{13}=0$, 
but it identifies this particular limit as the result of a 
well-defined symmetry.
Of course CP violation is not observable in oscillations, but it
can appear in neutrinoless $2\beta$ decay, since $m_i$ are in general complex
parameters.

One can ask the question if it is crucial for the above scenario 
to use Higgs triplets
$\xi_i$ (type II seesaw) instead of right-handed neutrinos $N_i$ 
(type I seesaw). 
In this last case the predictions depend on the source of the $N_i$ 
Majorana masses.
For definiteness, one can assume this source to be given by Higgs 
singlets $S_i$
which acquire super-heavy vacuum expectation values.
In order to reproduce as closely as possible the above pattern, 
let us make the following assignments:
\begin{equation}
N_i \sim (+,-),~(-,+),~(-,-), ~~~~ S_1 \sim (+,+),~S_2 \sim (-,-) ~.
\label{3nt2}\end{equation}
Then the neutrino mass matrix is given by
\begin{equation}
{\cal M}_\nu = - {\cal M}_D {\cal M}_R^{-1} {\cal M}_D^T =
- \left(\begin{array}{ccc}
a_\nu & 0 & 0 \\ 0 & b_\nu & d_\nu \\ 0 & e_\nu & c_\nu 
\end{array}\right)
\left(\begin{array}{ccc}
A_R & D_R & 0 \\ D_R & B_R & 0 \\ 0 & 0 & C_R 
\end{array}\right) ^{\textstyle{-1}}
\left(\begin{array}{ccc}
a_\nu & 0 & 0 \\ 0 & b_\nu & e_\nu \\ 0 & d_\nu & c_\nu 
\end{array}\right) 
\end{equation}
(a general method to obtain texture zeros in type I seesaw matrices using
flavor symmetries can be found in \cite{manyzero}).
In this case the diagonalization of ${\cal M}_D$ requires also a right-handed 
rotation, analogously to Eq.~(\ref{defi}). Therefore a non-zero $\theta_{13}$ 
is in general induced.  However, an interesting physical limit exists, such 
that $\theta_{13}$ is maintained to be zero, i.e. $M_3 \equiv C_R 
\rightarrow \infty$, so that the heaviest $N_3$ decouples, then it is easy 
to check that $\theta_{13}\rightarrow 0$ and, at the same time, 
$m_3\rightarrow 0$. The smallness of $\theta_{13}$ is now related to the 
inverted hierarchy of the spectrum.  Alternatively, we can simply eliminate 
$N_3$ from the beginning, i.e. keep only two right-handed neutrinos as in 
Ref.~\cite{gl04}, which obtained the same result using a $U(1)$ flavor 
symmetry.  In this scenario, since $\phi_2$ contributes both to ${\cal M}_l$ 
and to ${\cal M}_D$, the observable left-handed $2-3$ mixing angle 
receives contributions from both $U_l$ and $U_\nu$.

There are only two other ways for $\theta_{13}$ to be zero in Eq.~(10).  If 
$b_\nu c_\nu - d_\nu e_\nu = 0$, then again $m_3 = 0$ as well as 
$\theta_{13} = 0$ (but without $M_3 \to \infty$: here one eigenvalue 
of ${\cal M}_D$ vanishes instead).  The third way is to have $b_\nu d_\nu 
+ c_\nu e_\nu = 0$ (which allows ${\cal M}_D$ of Eq.~(10) to be diagonalized 
by a unitary transformation $U_L$ on the left and $U_R=1$ on the right), 
then $m_3$ remains arbitrary as in the model with Higgs triplets.  This form 
of the $2-3$ submatrix of ${\cal M}_D$ by itself is maintained for example
by the discrete symmetry $S_3$ \cite{cfm04}.

To test our model, we consider the details of the Higgs sector. Since the 
Higgs triplets are assumed to be very heavy, at the electroweak scale only 
the Higgs doublets are observable.  Using Eq.~(8), the Yukawa couplings of 
$\phi_1$ and $\phi_2$ are easily obtained as functions of $m_\mu$, $m_\tau$, 
$\theta_L$ and $\theta_R$, together with $v_1$ and $v_2$ subject to the 
constraint $\sqrt {v_1^2 + v_2^2} = 174$ GeV.  
The structure of Eq.~(\ref{defi}) 
tells us that leptonic flavors only change between $\mu$ and $\tau$, apart from
effects suppressed by the neutrino masses.  
In particular, the severe experimental constraints on $\mu\rightarrow e\gamma$
are automatically satisfied, as in the Standard Model. The 
couplings of $\phi^0_{1,2}$ are listed in Table 1.  The physical charged 
Higgs boson is given by
\begin{equation}
h^- = {v_2 \phi^-_1 - v_1 \phi^-_2 \over \sqrt {v_1^2 + v_2^2}}~,
\end{equation}
where $\phi^-_{1,2}$ couple to leptons as in Table 1, with $\mu$ replaced 
by $\nu_\mu$ and $\tau$ by $\nu_\tau$ respectively.

\begin{table}[htb]
\caption{Yukawa couplings of $\phi^0_{1,2}$.}
\begin{center}
\begin{tabular}{|c|c|c|c|c|}
\hline
 & $\phi^0_1 m_\mu/v_1$ & $\phi^0_1 m_\tau/v_1$ & $\phi^0_2 m_\mu/v_2$ & 
$\phi^0_2 m_\tau/v_2$ \\ 
\hline
$\mu \mu^c$ & $c_L^2 c_R^2 + s_L^2 s_R^2$ & $2 s_L c_L s_R c_R$ & $c_L^2 s_R^2 
+ s_L^2 c_R^2$ & $-2 s_L c_L s_R c_R$ \\
$\mu \tau^c$ & $(c_L^2 - s_L^2) s_R c_R$ & $-s_L c_L (c_R^2 - s_R^2)$ & 
$-(c_L^2 - s_L^2) s_R c_R$ & $s_L c_L (c_R^2 - s_R^2)$ \\
$\tau \mu^c$ & $s_L c_L (c_R^2 - s_R^2)$ & $-(c_L^2 - s_L^2) s_R c_R$ & $-s_L 
c_L (c_R^2 - s_R^2)$ & $(c_L^2 - s_L^2) s_R c_R$ \\
$\tau \tau^c$ & $2s_L c_L s_R c_R$ & $c_L^2 c_R^2 + s_L^2 s_R^2$ & $-2s_L c_L 
s_R c_R$ & $c_L^2 s_R^2 + s_L^2 c_R^2$ \\
\hline
\end{tabular}
\end{center}
\end{table}

In the case of ${\cal M}_\nu$ generated by Higgs triplets as in 
Eq.~(\ref{mnu}), we have 
$\theta_L = \theta_{23}$.  In the limit $\theta_L=\theta_{23}=\pi/4$ (which 
is preferred by the data) and neglecting $m_\mu$ versus $m_\tau$, the 
coupling of $h^-$ to leptons is given by
\begin{equation}
{m_\tau \over \sin 2 \beta \sqrt {v_1^2+v_2^2}} h^-  
\left[ 
  \sin 2 \theta_R \bar{\mu} \left( {1 - \gamma_5 \over 2} \right) \nu_\mu 
- \cos 2 \theta_R \bar{\tau}\left( {1 - \gamma_5 \over 2} \right) \nu_\mu 
- \cos 2 \beta    \bar{\tau}\left( {1 - \gamma_5 \over 2} \right) \nu_\tau 
\right] + {\rm h.c.} ~,
\label{hmy}
\end{equation}
where $\tan \beta = v_2/v_1$.  This implies that
\begin{equation}
{\Gamma (h^- \to \mu^- \nu) \over \Gamma (h^- \to \tau^- \nu)} = 
{\sin^2 2 \theta_R \over \cos^2 2 \beta + \cos^2 2 \theta_R}~,
\end{equation}
instead of $m_\mu^2/m_\tau^2$, as in the usual (MSSM like) two Higgs 
doublet models.  
Thus this ratio is, in general, not suppressed and is a good experimental 
test of this model.

The neutral Higgs boson of the Standard Model (with the usual Yukawa 
couplings to leptons) is
\begin{equation}
H^0 = {\sqrt 2 (v_1 \Re e\phi^0_1 + v_2 \Re e\phi^0_2) \over \sqrt 
{v_1^2 + v_2^2}}~,
\end{equation}
but it is not in general a mass eigenstate in a two-Higgs-doublet model.  
It mixes with 
\begin{equation}
h^0 = {\sqrt 2 (v_2 \Re e\phi^0_1 - v_1 \Re e\phi^0_2) \over \sqrt 
{v_1^2 + v_2^2}}~,
\end{equation}
which couples to leptons, in the same limit as in Eq.~(\ref{hmy}), 
according to
\begin{equation}
{m_\tau \over \sqrt 2 \sin 2 \beta \sqrt {v_1^2 + v_2^2}} 
h^0 \left[\sin 2 \theta_R 
\bar \mu \mu - \cos 2 \theta_R \bar \tau \left( {1-\gamma_5 \over 2} 
\right) \mu - \cos 2 \theta_R \bar \mu \left( {1+\gamma_5 \over 2} \right) 
\tau - \cos 2 \beta \bar \tau \tau \right]~.
\label{hzy}
\end{equation}
In general, $H^0$ may also mix with 
\begin{equation}    
A^0 = {\sqrt 2 (v_2 \Im m\phi^0_1 - v_1 \Im m\phi^0_2) \over \sqrt 
{v_1^2 + v_2^2}}
\end{equation}
which couples to leptons according to
\begin{equation}
{-i m_\tau \over \sqrt 2 \sin 2 \beta \sqrt {v_1^2 + v_2^2}} 
A^0 \left[\sin 2 \theta_R 
\bar \mu \gamma_5 \mu + \cos 2 \theta_R \bar \tau \left( {1-\gamma_5 \over 2} 
\right) \mu - \cos 2 \theta_R \bar \mu \left( {1+\gamma_5 \over 2} \right) 
\tau - \cos 2 \beta \bar \tau \gamma_5 \tau \right]~.
\label{azy}
\end{equation}

If the Higgs potential has exact $Z_2 \times Z_2$ symmetry, then one can 
check that $CP$ is 
conserved and $A^0$ is a mass eigenstate (with odd $CP$) and does not mix 
with $h^0$ and $H^0$ which are even under $CP$.  The decay of $A^0$ is thus 
another distinct signature of this model: the branching fractions of $A$ to 
$\tau^+ \tau^-$, $\tau^+ \mu^- + \mu^+ \tau^-$, and $\mu^+ \mu^-$ are 
proportional to $\cos^2 2 \beta$, $\cos^2 2 \theta_R$, and $\sin^2 2 
\theta_R$ respectively.  If $Z_2 \times Z_2$ is allowed to be broken by 
soft terms of the Higgs potential, then $CP$ is violated and all 3 
neutral Higgs bosons $A^0, h^0, H^0$ mix with one another.  In the 
following we assume, for simplicity, that $A^0$ is a mass eigenstate and that
\begin{equation}
\left(\begin{array}{c} h_1^0 \\ h_2^0 \end{array}\right)
=
\left(\begin{array}{cc} \cos\alpha & \sin\alpha \\ -\sin\alpha & 
\cos\alpha \end{array}\right)
\left(\begin{array}{c} H^0 \\ h^0 \end{array}\right) ~,
\end{equation}
where $h^0_{1,2}$ are the eigenstates with masses $m_{1,2}$.

Because of Eqs.~(\ref{hmy}), (\ref{hzy}) and (\ref{azy}), the 
flavor-changing processes $\tau \to 
\mu \mu \mu$ and $\tau \to \mu \gamma$ are predicted, as well as an 
additional contribution to the muon anomalous magnetic moment.  Consider 
first $\tau \to 3\mu$.  It proceeds through $A^0$ and $h^0$ exchange. 
Although $h^0$ mixes with $H^0$, the latter does not couple to $\bar \mu 
\tau$ and its coupling to $\bar \mu \mu$ is proportional to $m_\mu$. 
We obtain
\begin{equation}
\Gamma(\tau \to 3\mu) = \left[ {m_\tau^2 \sin 2 \theta_R \cos 2 \theta_R 
\over 2 \sin^2 2 \beta (v_1^2 + v_2^2)} \right]^2 {m_\tau^5 \over 4096 
\pi^3} \left( {1 \over m_A^4} + {1 \over  m_{h^0}^4} + 
{2 \over 3 m_{A}^2  m_{h^0}^2}\right),
\end{equation}
where $ m_{h^0}$ is the effective contribution of $h^0$ exchange:
\begin{equation}
\frac{1}{ m_{h^0}^2}\equiv \frac{\sin^2\alpha}{m^2_1}
+\frac{\cos^2\alpha}{m^2_2} ~.
\end{equation}
Numerically, for $m_A =  m_{h^0} = 100$ GeV and $\sin 2 \theta_R 
\cos 2 \theta_R/\sin^2 2 \beta = 1$, this implies a branching fraction of 
$4.5 \times 10^{-9}$, well below the present experimental upper bound 
\cite{pdg} of $1.9 \times 10^{-6}$.

In the same approximation as above, the radiative decay rate of $\tau \to \mu 
\gamma$ is given by
\begin{equation}
\Gamma (\tau \to \mu \gamma) = {\alpha_{em} m_\tau^5 \over (64 \pi^2)^2} 
(|A_L|^2 + |A_R|^2),
\end{equation}
where
\begin{equation}
A_L = {1 \over 3} \left[ {m_\tau^2 \sin 2 \theta_R \cos 2 \theta_R \over 
2 \sin^2 2 \beta (v_1^2 + v_2^2)} \right] \left[ {1 \over m_A^2} + {1 \over 
m_{h^0}^2} - {1 \over m_{h^-}^2} \right],
\end{equation}
and
\begin{equation} 
A_R = \left[ {m_\tau^2 \cos 2 \theta_R \cos 2 \beta \over 2 \sin^2 2 \beta 
(v_1^2 + v_2^2)} \right] \left[ \sum_{i=1}^{2}{k_i \over m_i^2} 
\left( {8 \over 3} + 
2 \ln {m_\tau^2 \over m_i^2} \right) - {1 \over m_A^2} \left( {10 \over 3} 
+ 2 \ln {m_\tau^2 \over m_A^2} \right) \right],
\end{equation}
with $k_1\equiv \sin^2\alpha-\sin\alpha\cos\alpha\tan2\beta$,
$k_2\equiv \cos^2\alpha+\sin\alpha\cos\alpha\tan2\beta$.
Numerically (using
$m_A=m_1=m_2=m_{h^-}=100$ GeV and $\cos 2 \theta_R =  
\sin 2 \theta_R = \sin 2 \beta = 1/\sqrt{2}$), this 
implies a branching fraction of $2.2 \times 10^{-12}$, 
again well below the experimental upper bound \cite{pdg} 
of $1.1 \times 10^{-6}$.

We computed also the contribution to the anomalous magnetic moment 
of the muon $a_\mu\equiv (g_\mu-2)/2$ 
from 1-loop diagrams mediated by $h^-$, $h^0$ and $A^0$:
\begin{eqnarray}
\delta a_\mu &=& \frac{m_\mu^2 m_\tau^2}{32\pi^2(v_1^2+v_2^2)\sin^22\beta}
\left\{
\cos^2 2\theta_R \left[ 
\frac{1}{3 m^2_{h^0}}+\frac{1}{3 m^2_{A}}
\right] + \right. \nonumber\\ &+&
\left.\sin^2 2\theta_R \left[ \sum_{i=1}^2
\frac{k_i}{m_i^2}\left(-\frac 73 -
2\log\frac{m_\mu^2}{m_i^2}\right) +
\frac{1}{m_{A}^2}\left(\frac{11}{3} +
2\log\frac{m_\mu^2}{m_{A}^2}\right)-\frac{1}{3 m^2_{h^-}}
\right]
\right\}~,
\end{eqnarray}
where $k_1\equiv \sin^2\alpha$, $k_2\equiv \cos^2\alpha$.
Using the parameter values given above, 
$\delta a_\mu \approx 6.2 \times 10^{-13}$,
that is much smaller than the present uncertainty 
($\sim 10^{-9}$) and
therefore negligible as a possible explanation of the discrepancy 
($\sim 3 \times 10^{-9}$) between the
Standard Model prediction \cite{hmnt} and the experimental value \cite{g-2}.

In general, the leptonic Yukawa couplings of this model are at most of order 
$m_\tau/M_W$ which is small enough to suppress all indirect 
Lepton-Flavor-Violation effects much below the present experimental 
upper bounds, unless $\tan\beta$ turns out to be very large. Thus the 
best hope of testing this model is through the direct production and decay 
of the extra Higgs bosons as already discussed.

Let us briefly review some features of the Majorana neutrino mass 
matrix with $\theta_{13} = 0$. In the basis where ${\cal M}_l$ is diagonal, 
Eqs.~(\ref{one}) and (\ref{U}) imply
\begin{equation}
{\cal M}_\nu = \pmatrix{c_{12}^2 m_1 + s_{12}^2 m_2 & 
s_{12} c_{12} c_{23} (m_1-m_2) & s_{12} c_{12} s_{23} (m_1-m_2) \cr 
s_{12} c_{12} c_{23} (m_1-m_2) & c_{23}^2 (s_{12}^2 m_1 + c_{12}^2 m_2) + 
s_{23}^2 m_3 & s_{23} c_{23} (s_{12}^2 m_1 + c_{12}^2 m_2 - m_3) \cr 
s_{12} c_{12} s_{23} (m_1-m_2) & s_{23} c_{23} (s_{12}^2 m_1 + c_{12}^2 m_2 
- m_3) & s_{23}^2 (s_{12}^2 m_1 + c_{12}^2 m_2) + c_{23}^2 m_3}.
\label{zerol}
\end{equation}
As shown in Ref.~\cite{gjklst}, this matrix by itself has a $Z_2$ symmetry.  
This may also be understood by its form invariance \cite{ma03}, i.e.
\begin{equation}
U {\cal M}_\nu U^T = {\cal M}_\nu ~,
\end{equation}
where
\begin{equation}
U = \pmatrix{1 & 0 & 0 \cr 0 & \cos 2 \theta_{23} & \sin 2 \theta_{23} \cr 
0 & \sin 2 \theta_{23} & -\cos 2 \theta_{23}}, ~~~ U^2 = 1~.
\end{equation}
The matrix of Eq.~(\ref{zerol}) was 
in fact obtained previously as the remnant of a complete $D_4 \times Z_2$ 
model \cite{gjklt}.  Another model \cite{fkmt} based on the quaternion 
group $Q_8$ also obtains this structure (if one CP phase is put to zero) 
with the further restriction
\begin{equation}
({\cal M}_\nu)_{23}=0 ~~~\Leftrightarrow ~~~ 
s_{12}^2 m_1 + c_{12}^2 m_2 = m_3~.
\end{equation}
If $c_{23} = s_{23}$ in Eq.~(\ref{zerol}), then ${\cal M}_\nu$ 
has the $Z_2$ symmetry 
proposed in Ref.~\cite{allp}, which is realized in the $A_4$ model \cite{a4}, 
with $m_1=m_2=-m_3$ (before radiative corrections).  
These examples and others in Ref. \cite{low} 
show that our present proposal of $Z_2 \times Z_2$ is 
not unique for obtaining $\theta_{13} = 0$, but is rather the simplest 
scenario and it is also consistent with arbitrary charged-lepton and Majorana 
neutrino masses. 
It should also be noted that after the heavy Higgs triplets 
(or the right-handed
neutrinos) are integrated 
away, the effective Lagrangian of this model (including the Higgs doublets) 
conserves $L_e$ and $L_\mu + L_\tau$ separately, broken only by the very 
small Majorana neutrino masses.

Quarks can be incorporated into this model, for example, assigning
\begin{equation}
(u_i,d_i),~u_i^c,~d_i^c \sim (-,-),~(-,+),~(+,-) ~.
\label{3nt3}\end{equation}
In this way both up and down quark mass matrices contain only $1-2$ mixing,
since the off-diagonal entries are induced by $\phi_2\sim(+,-)$. 
Therefore the Cabibbo
mixing can be reproduced while other mixing angles are suppressed.
The three generations of fermions in Eqs.(\ref{3nt1}),
(\ref{3nt2}) and (\ref{3nt3}) are associated with the 
nontrivial representations of $Z_2 \times Z_2$. They
can be identified as the three 
components of the corresponding triplets of $SO(3)$, which breaks 
down to $Z_2 \times Z_2$ (i.e. a rectangle embedded inside a sphere).
An alternative way to incorporate quarks in the model is to extend the 
discrete symmetry to $Z_2 \times Z_2 \times Z_2$, with leptons transforming 
trivially under the third $Z_2$ and quarks trivially under the second $Z_2$. 
The addition of $\Phi_3 \sim (-,+,-)$ and $\Phi_4 \sim (+,+,-)$ would then 
generate the complete quark mixing matrix, as in the $Q_8$ model \cite{fkmt}.

Since left-handed and right-handed fermions transform in the same way under
$Z_2\times Z_2$, one could embed this model in a left-right symmetric theory.
In particular, theories based on $SO(10)$ are a natural framework to provide 
both types of seesaw mechanism, since their particle spectrum may include both
super-heavy right-handed neutrinos $N_i$ and scalar isotriplets 
$\xi_i$.

In conclusion, we pointed out that the absence of $1-3$ mixing in the 
lepton sector can be explained if the Standard Model Lagrangian is extended
to include two Higgs doublets and an appropriate 
source for neutrino Majorana masses,
in such a way to respect a $Z_2\times Z_2$ family symmetry.
In this scenario the atmospheric mixing angle originates in the $\mu-\tau$
sector of the charged lepton mass matrix and it relates with
predictable Lepton-Flavor-Violation effects: the physical Higgs bosons
have specific decay rates into muons and taus, while their indirect
contributions to $\tau\rightarrow \mu\gamma$, $\tau\rightarrow 3\mu$ and
$g_\mu-2$ are in general negligible.
The solar mixing angle originates in the neutrino mass matrix, that can be
generated either by two Higgs triplets or by two right-handed neutrinos.

This work was supported in part by the U.~S.~Department of Energy
under Grant No.~DE-FG03-94ER40837.

\bibliographystyle{unsrt}

\begin{thebibliography}{99}

\bibitem{review} For a recent review, see for example K. Heeger, Talk at 
Seesaw25, Paris, (June 2004),
http://seesaw25.in2p3.fr/trans/heeger.pdf .

\bibitem{valencia}
M.~Maltoni, T.~Schwetz, M.~A.~Tortola and J.~W.~F.~Valle,
New J. Phys. {\bf 6} (2004) 122 [hep-ph/0405172].

\bibitem{vibe}
F.~Vissani, M.~Narayan and V.~Berezinsky,
Phys.\ Lett.\ B {\bf 571} (2003) 209
[hep-ph/0305233].

\bibitem{nufa}
C.~Albright {\it et al.}  [Neutrino Factory/Muon Collider Collaboration],
physics/0411123.

\bibitem{frpe}
P.~H.~Frampton, S.~T.~Petcov and W.~Rodejohann,
Nucl.\ Phys.\ B {\bf 687} (2004) 31
[hep-ph/0401206].

\bibitem{alfe}
G.~Altarelli, F.~Feruglio and I.~Masina,
Nucl.\ Phys.\ B {\bf 689} (2004) 157
[hep-ph/0402155].

\bibitem{roma}
A.~Romanino,
Phys.\ Rev.\ D {\bf 70} (2004) 013003
[hep-ph/0402258].

\bibitem{anki}
S.~Antusch and S.~F.~King,
Phys.\ Lett.\ B {\bf 591} (2004) 104
[hep-ph/0403053].

\bibitem{raid}
M.~Raidal,
Phys.\ Rev.\ Lett.\  {\bf 93} (2004) 161801
[hep-ph/0404046].

\bibitem{feru}
F.~Feruglio,
hep-ph/0410131.

\bibitem{mism}
H.~Minakata and A.~Y.~Smirnov,
hep-ph/0405088.

\bibitem{grla}
W.~Grimus and L.~Lavoura,
JHEP {\bf 0107} (2001) 045
[hep-ph/0105212].

\bibitem{low} C. I. Low, Phys. Rev. D {\bf 70} (2004) 073013
[hep-ph/0404017].


\bibitem{gjklt} W. Grimus, A. S. Joshipura, S. Kaneko, L. Lavoura, 
and M. Tanimoto, JHEP {\bf 0407} (2004) 078 [hep-ph/0407112].

\bibitem{gjklst} W. Grimus, A. S. Joshipura, S. Kaneko, L. Lavoura, 
H. Sawanaka, and M. Tanimoto, hep-ph/0408123. 

\bibitem{ms98} E. Ma and U. Sarkar, Phys. Rev. Lett. {\bf 80} (1998) 5716; 
E. Ma, Phys. Rev. Lett. {\bf 81} (1998) 1171.

\bibitem{manyzero} W. Grimus, A. S. Joshipura, L. Lavoura, 
and M. Tanimoto, Eur. Phys. J. C {\bf 36} (2004) 227 [hep-ph/0405016].

\bibitem{gl04} W. Grimus and L. Lavoura, hep-ph/0410279.

\bibitem{cfm04} S.-L. Chen, M. Frigerio, and E. Ma, Phys. Rev. D {\bf 70} 
(2004) 073008 [hep-ph/0404084].

\bibitem{pdg} Particle Data Group: S. Eidelman {\it et al.}, Phys. Lett. B 
{\bf 592} (2004) 1.

\bibitem{hmnt} See for example K. Hagiwara {\it et al.}, 
Phys. Rev. D {\bf 69} (2004) 093003 [hep-ph/0312250].

\bibitem{g-2} G. W. Bennett {\it et al.}, Muon (g-2) Collaboration, Phys. 
Rev. Lett. {\bf 89} (2002) 101804.

\bibitem{ma03} E. Ma, Phys. Rev. Lett. {\bf 90} (2003) 221802.

\bibitem{fkmt} M. Frigerio, S. Kaneko, E. Ma, and M. Tanimoto, hep-ph/0409187.

\bibitem{allp} E. Ma, Phys. Rev. D {\bf 66} (2002) 117301.

\bibitem{a4} E. Ma and G. Rajasekaran, Phys. Rev. D {\bf 64} (2001) 113012; 
K. S. Babu, E. Ma, and J. W. F. Valle, Phys. Lett. B {\bf 552} (2003) 207.

\end{thebibliography}

\end{document}